\documentclass[aps,preprint,showpacs,floats,epsf,epsfig,nofootinbib,12pt]{revtex4}
\usepackage{graphicx}
\usepackage{epsfig}
\usepackage[dvips]{color}
\usepackage{subfigure}
\usepackage{multirow}

\def\beq{\begin{eqnarray}}
\def\eeq{\end{eqnarray}}

\begin{document}

\title{Tetraquark state $X(6900)$ and the interaction between diquark and antidiquark}

\vspace{1cm}

\author{ Hong-Wei Ke$^1$\footnote{khw020056@tju.edu.cn}, Xin Han$^1$, Xiao-Hai
Liu$^1$\footnote{xiaohai.liu@tju.edu.cn} and Yan-Liang Shi
        $^2$\footnote{shi@cshl.edu}  }

\affiliation{  $^{1}$ School of Science, Tianjin University, Tianjin 300072, China \\
  $^{2}$ Cold Spring Harbor Laboratory, Cold Spring Harbor, NY 11724, USA }

\vspace{12cm}

\begin{abstract}

Recently LHCb declared a new structure $X(6900)$ in the final
state di-$J/\psi$ which is popularly regarded as a $cc$-$\bar
c\bar c$ tetraquark state. Within the Bethe-Salpeter (B-S)
framework we study the possible $cc$-$\bar c\bar c$
 bound states and the interaction between diquark ($cc$) and antidiquark ($\bar c\bar c$). In this work $cc$ ($\bar c\bar c$) is treated as a color anti-triplet (triplet) axial-vector so the quantum numbers of $cc$-$\bar c\bar c$ bound state
 are $0^+$, $1^+$ and $2^+$. Learning from the interaction in meson case and using the effective coupling we suggest the interaction kernel for the diquark and antidiquark system. Then we deduce the B-S equations for different quantum numbers. Solving these equations numerically we find the spectra of some excited states can be close to the mass of
$X(6900)$  when we assign appropriate values for parameter
$\kappa$ introduced in the interaction (kernel). We also briefly
calculate the spectra of $bb$-$\bar b\bar b$
 bound states.
Future measurement of  $bb$-$\bar b\bar b$ state  will help us to
determine the exact form of effective interaction.

\pacs{12.39.Mk, 12.40.-y ,14.40.Nd}

\end{abstract}

\maketitle

\section{Introduction}
Not long ago, LHCb declared a narrow structure about 6.9 GeV (
named as $X(6900)$ or $T_{cc\bar c\bar c}$) and a broad structure
about twice the $J/\psi$ mass  in the final state
di-$J/\psi$\cite{Aaij2020}. Based on no-interference fit, the new
state's mass and width are $6905\pm 11\pm 7$ MeV and $80\pm 19\pm
33$ MeV, while based on the simple model with interference they
are $6886\pm 11\pm 11$ MeV and $168\pm 33\pm 69$ MeV. In the past
twenty years many exotic states named $X,\,Y,\,Z$ have been
experimentally observed. However, $X(6900)$ is a novel state,
since it seems a fully-charmed multi-quark state. This new
discovery has inspired many theoretical
interests\cite{Cao:2020gul,Guo:2020pvt,Zhu:2020xni,Gordillo:2020sgc,Ma:2020kwb,Deng:2020iqw,
Zhao:2020cfi,Wang:2020ols,Chen:2020xwe,Albuquerque:2020hio,Karliner:2020dta,Faustov:2020qfm,Zhao:2020nwy,Giron:2020wpx,Jin:2020jfc,
Lu:2020cns,Zhang:2020xtb,Wan:2020fsk,Yang:2020wkh,Gong:2020bmg,Zhu:2020snb,Dong:2020nwy}.
Many authors suggest that the new resonance $T_{cc\bar c\bar c}$
can be a $cc$-$\bar c\bar c$ tetraquark
state\cite{Zhang:2020xtb,Faustov:2020qfm,Zhao:2020nwy,Karliner:2020dta,Giron:2020wpx,Wang:2020ols,Jin:2020jfc,Lu:2020cns,Deng:2020iqw},
a $cc$-G-$\bar c\bar c$ hybrid\cite{Wan:2020fsk}, a $c\bar c_{\rm
octet}$-$ c\bar c_{\rm octet}$ tetraquark\cite{Yang:2020wkh}, a
dynamically generated resonance pole\cite{Gong:2020bmg} or a light
Higgs-like boson\cite{Zhu:2020snb}. Few people believe $T_{cc\bar
c\bar c}$ is a molecular state because the interaction between two
charmonia is too weak to form a molecular state.

In fact some
authors\cite{Barnea:2006sd,Berezhnoy:2011xn,Bedolla:2019zwg,Debastiani:2017msn,Wang:2018poa}
had been explored the $cc$-$\bar c\bar c$ bound state before the
measurement of LHCb. In this paper we will study the possible
bound states between the diquark ($cc$) and antidiquark ($\bar
c\bar c$) using the Bethe-Salpeter (B-S) equation which is a
relativistic two-body bound state equation. We want to explore the
interaction between diquark and antidiquark by this study. As we
know, if the orbital angular momentum is zero, two $c$  quarks
 constitute an axial-vector diquark, but it may be a color
anti-triple state or sextet one.  The one-gluon exchange potential
is attractive for anti-triple state but repulsive for sextet one.
Therefore, in this work we will only consider the axial-vector
diquark (antidiquark) of color anti-triplet (color triplet). In
quantum field theory, at the tree level, two particles interact
with each other by exchanging an intermediary messenger particle.
For the present case, the tetraquark state consists of one diquark
and one antidiquark. Apparently $cc$ in $\bar 3_c$ presentation
and $\bar c\bar c$ in $3_c$ one can be regards as an antiquark and a
quark respectively, so gluons are exchanged between $cc$ and $\bar
c\bar c$. And one can employ a phenomenological potential similar
to the meson case to study the diquark and antidiquark bound
state. However, since diquark isn't the fundamental particle,  the
coupling between diquark and gluon doesn't exist in QCD
Lagrangian. In Ref.\cite{Kroll:1990hg,Korner:1992uw} the authors
suggested an effective coupling between diquark and gluon, which
is different from that between quark and gluon so one cannot use
the phenomenological potential for the mesons (Cornell
potential\cite{Eichten:1978tg,Eichten:1979ms}) directly.
Therefore, we have to modify the form of potential between diquark and
antidiquark. With that coupling in Ref.\cite{Korner:1992uw} we can
modify the coulomb part in the potential, but we still lack a
proper treatment for the confine part of the potential. To address
this issue, we follow the approach in
Ref.\cite{Guo:1998ef,Guo:1996jj} and introduce a parameter
$\kappa$ for the confine part in the Cornell potential.

Initially B-S equation was used to explore the bound state of two
fermions. Later some authors extended this approach to study the
bound state composed of one fermion and one
boson\cite{Guo:1998ef,Weng:2010rb,Li:2019ekr}.  In
Ref.\cite{Guo:2007mm,Feng:2011zzb} the authors employed B-S
equation to study the $K\bar K$ and $BK$ molecular stated and
their decays. Later we extended the method to explore some other
systems\cite{Ke:2012gm,Ke:2019bkf}. Diquark and antidiquark bound
state is a novel system in this approach and the effective
interaction hasn't been studied in depth due to the lack of data.
In this work, we will study the phenomenological potential between diquark and antidiquark, and establish the formalism of effective interaction for this bound state system.

Here we only concern the state where the orbital angular momentum
between the two constituents ($cc$ and $\bar c\bar c$) is zero
($l=0)$ so the $J^{P}$ of the molecular state can be $0^+$, $1^+$
or $2^+$. Following Ref.\cite{Guo:1998ef}, we explore the
potential between $cc$ and $\bar c\bar c$ and then deduce the B-S
equations for different quantum numbers. By solving these B-S
equations numerically we obtain the spectra and B-S wave function
of the $cc$-$\bar c\bar c$ bound states. We use experimental data
to determine the free parameters in the expression of the
effective potential.

This paper is organized as follows. In section II we deduce the B-S equations for the
$0^+$, $1^+$ and $2^+$ diquark-antidiquark states. Then in section III we present
our numerical results and explicitly display all input
parameters. Section IV is devoted to the summary and discussion. As
indicated before, in this work we concentrate on the case of
$cc$-$\bar c\bar c$ states, but we also
briefly calculate the spectra of the $bb$-$\bar b\bar b$ states.

\section{The bound states of diquark and antidiquark }

\begin{figure}
\begin{center}
\begin{tabular}{ccc}
\scalebox{0.8}{\includegraphics{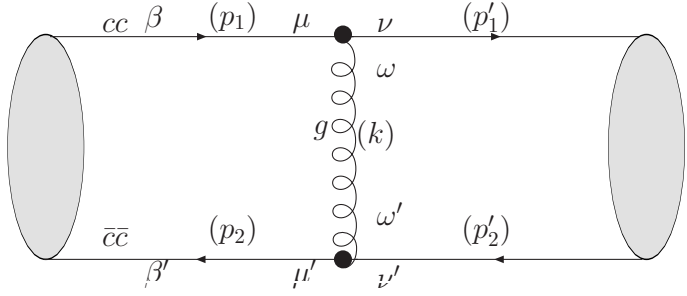}}
\end{tabular}
\end{center}
\caption{the tetraquark state of $cc$ and $\bar c\bar
c$}\label{bound}
\end{figure}

A diquark ($cc$) and an antidiquark ($\bar c\bar c$) can interact
by exchanging gluon, and the scenario is depicted in
Fig.\ref{bound}. Since $cc$ ($\bar c\bar c$) consists of two same
quarks (antiquark), the ground state of $cc$ ($\bar c\bar c$)
possesses $1^+$ quantum number.

The relative and total momenta of the bound state in the equations
are defined as
\begin{eqnarray} p = \eta_2p_1 -
\eta_1p_2\,,\quad p' = \eta_2p'_1 - \eta_1p'_2\,,\quad P = p_1 +
p_2 = p'_1 + p'_2 \,, \label{momentum-transform1}
\end{eqnarray}
where $p$ and $p'$ are the relative momenta before and after the effective vertices,
$p_1$ ($p'_1$) and $p_2$ ($p'_2$) are the momenta of the
constituents before  and after the effective vertices, $P$ is the total
momentum of the bound state, $\eta_i = m_i/(m_1+m_2)$ and $m_i\,
(i=1,2)$ is the mass of the $i$-th constituent (diquark or antidiquark).

\subsection{The B-S equation of $0^+$ state which is composed of two axial-vectors}
The B-S wave function of $0^+$  state composed of two axial-vector can be defined as
\begin{eqnarray}\label{definition-BS}  \langle 0 | {\rm
T}\,\phi_{1}^\alpha(x_1)\phi^{\alpha'}_2(x_2) | \mathcal{S} \rangle={\chi}_{{}_\mathcal{S}}^{}(x_1,x_2)g^{\alpha\alpha'}\,.
\end{eqnarray}
The corresponding B-S equation can be written as
\begin{eqnarray} \label{4-dim-BS21}
\chi_{_{{}_\mathcal{S}}}({
p})g_{\beta\beta'}=\Delta_{1\mu\beta}\int{d^4{p}'\over(2\pi)^4}\,K^{\mu\mu'\nu\nu'}({
p},{ p}')\chi_{_{{}_\mathcal{S}}}^{}({
p}')\Delta_{2\mu'\beta'}g_{\nu\nu'},
\end{eqnarray}
{where the propagators
$\Delta_{1\mu\beta}=(-g_{\mu\beta}+p_{1\mu}p_{1\beta}/{m_1}^2)/({p_1}^2-{m_1}^2+i\epsilon)$
and
$\Delta_{2\mu'\beta'}=(-g_{\mu\beta}+p_{2\mu'}p_{2\beta'}/{m_2}^2)/({p_2}^2-{m_2}^2+ i\epsilon)$}.

Recall that the kernel between a quark and an antiquark\cite{Chang:2010kj} has the following form:
\begin{eqnarray}\label{mkernel}
K({ p},{ p}')=K_v({ p},{ p}')\gamma^\mu\bigotimes\gamma_\mu+ K_s({
p},{ p}')I\bigotimes I \ ,
\end{eqnarray}
where $K_v({ p},{ p}')$ is called vector potential which comes
from single gluon exchange, $K_s({ p},{ p}')$ is scalar potential
which is responsible for the confinement and after taking instantaneous approximation $p_0'=p_0$
\begin{eqnarray}
K_v(\mathbf{q})=-\frac{16\pi\alpha_s}{3(\mathbf{q}^2+\alpha^2)},\,\,K_s(\mathbf{q})=-\frac{8\pi\lambda}{(\mathbf{q}^2+\alpha^2)^2}
+(2\pi)^3(\frac{\lambda}{\alpha}+V_0)\delta^3(\mathbf{q}) \  ,
\end{eqnarray}
{where $q$ is equal to $p-p'$
($\mathbf{q}$ is its three-momentum)}, $V_0$ is the zero-point energy
term and $\alpha$ is a very small parameter. Since a diquark is in
$\bar 3_c$ presentation and an antiquark is in $3_c$ presentation,
the interaction between diquark and antidiquark should be similar
to that between quark and antiquark. However diquark and
antidiquark are not the fundamental particles in standard model,
so effective interaction vertex of the vector diquark-gluon
coupling can be parameterized as
\begin{eqnarray}
-ig_s t^a[(p_1+{p'}_1)^\omega g^{\mu\nu}-k_vp_1^\mu
g^{\omega\nu}-(1+k_v){p}_1^\nu g^{\mu\omega}-k_v{p'}_1^\nu g^{\mu\omega}-(1+k_v){p'}_1^\mu
g^{\omega\nu}]F_V(q^2),
\end{eqnarray}
{where $g_s$ is strong coupling
constant, $t^a$ is color matrix, and $k_v$ is the anomalous magnetic momentum of
vector diquark. In the non-relativistic quark model, $k_v$ should be $1$. The form factor $ F_V(q^2)$ is parametrized as $[Q_1^2/(Q_1^2 -q^2)]^2$ where $Q^2_1$ is a parameter which freezes
$F_V(q^2)$ when $q^2$ is small. In Ref.\cite{Kroll:1990hg} the authors fixed
$Q^2_1$ and $k_v$ to be 1.58 GeV$^2$ and 1.16 by fitting the diquark model of nucleon  based on the data of electron-proton
and electron-neutron
cross-sections\cite{Arnold:1986nq,Rock:1982gf}.}

If only single gluon exchange is accounted,
the kernel can be
\begin{eqnarray} \label{kernel}
&&-\frac{16\pi\alpha_s}{3}[F_V(q^2)]^2[(p_1+{p'}_1)^\omega
g^{\mu\nu}-k_vp_1^\mu g^{\omega\nu}-(1+k_v){p}_1^\nu
g^{\mu\omega}-k_v{p'}_1^\nu g^{\mu\omega}-(1+k_v){p'}_1^\mu
g^{\omega\nu}]\nonumber\\&&[(p_2+{p'}_2)^{\omega'}
g^{\mu'\nu'}-k_vp_2^{\mu'} g^{\omega'\nu'}-(1+k_v){p}_2^{\nu'}
g^{\mu'\omega'}-k_v{p'}_2^{\nu'}
g^{\mu'\omega'}-(1+k_v){p'}_2^{\mu'}
g^{\omega'\nu'}]\frac{-ig_{\omega\omega'}}{q^2}.
\end{eqnarray}
Compared with the kernel in Eq. (\ref{mkernel}) we have
$$K^{\mu\mu'\nu\nu'}({ p},{
p}')=K_{v}(p,p')\Gamma^{\omega\mu\nu}_1\bigotimes\Gamma^{\omega'\mu'\nu'}_2[F_V(q^2)]^2(-ig_{\omega\omega'})+
K_s({ p},{ p}')I\bigotimes Ig^{\mu\mu'}g^{\nu\nu'}\kappa$$ where
$\Gamma^{\omega\mu\nu}_1=[(p_1+{p'}_1)^\omega g^{\mu\nu}-k_vp_1^\mu
g^{\omega\nu}-(1+k_v){p}_1^\nu g^{\mu\omega}-k_v{p'}_1^\nu g^{\mu\omega}-(1+k_v){p'}_1^\mu
g^{\omega\nu}]$,
$\Gamma^{\omega'\mu'\nu'}_2=[(p_2+{p'}_2)^{\omega'} g^{\mu'\nu'}-k_vp_2^{\mu'}
g^{\omega'\nu'}-(1+k_v){p}_2^{\nu'} g^{\mu'\omega'}-k_v{p'}_2^{\nu'} g^{\mu'\omega'}-(1+k_v){p'}_2^{\mu'}
g^{\omega'\nu'}]$ and $\kappa$ is a parameter introduced to compensate the dimension of the second term and its dimension is the quadratic of mass.

We need time $g_{\beta\beta'}$ on both sides in Eq.
(\ref{4-dim-BS21}) and sum the same indexes. In order to achieve
the three-dimensional Salpeter we take the instantaneous
approximation $p'_0=p_0$ in the kernel and perform the integral
over $p_0$ on both side of Eq. (\ref{4-dim-BS21}). The right hand
side of expression is a contour integral in the complex place. By
choosing a proper contour\cite{Guo:2007mm,Feng:2011zzb} one can
obtain the corresponding B-S equation in three-momentum space
\begin{eqnarray} \label{3-dim-BS11}
\mathcal{\psi}_{{}_\mathcal{S}}^{}({\bf |p|}) =&&\frac{
1}{8E_1E_2[E^2-(E_1+E_2)^2](E+E_1-E_2)}\int{d^3\mathbf{p}'\over(2\pi)^3}\,
\psi_{{}_\mathcal{S}}^{}({\bf
|p|}')\nonumber\\&&\{E_2(E-E_1-E_2)[C_{\mathcal{S}1}|_{_{p_0=-E/2-E_1}}K_{v}(\mathbf{q})|F_V(\mathbf{q}^2)|^2+C_{\mathcal{S}2}|_{_{p_0=-E/2-E_1}}K_{s}(\mathbf{q})\kappa]\nonumber\\&&
+E_1(E+E_1+E_2)[C_{\mathcal{S}1}|_{_{p_0=E/2-E_2}}K_{v}(\mathbf{q})|F_V(\mathbf{q}^2)|^2+C_{\mathcal{S}2}|_{_{p_0=E/2-E_2}}K_{s}(\mathbf{q})\kappa]\}
\,,
\end{eqnarray}
where $E$ is the total energy of the bound state, $E_i =
\sqrt{{\bf p}^2 + m_i^2}$ and $\psi_{{}_\mathcal{S}}^{}({\bf
|p|})=\int{d{p_0}\over(2\pi)}\chi_{_{{}_\mathcal{S}}}^{}({ p})$ is
the B-S wave function in the three-momentum space. The explicit
expressions of $C_{\mathcal{S}1}$ and $C_{\mathcal{S}2}$ are
presented in Appendix.

\subsection{The B-S equation of $1^+$ state which is composed of two vectors}

 The B-S wave function of $1^+$ state composed of two axial-vectors is defined
 \begin{eqnarray} \label{4-dim-BS22}
\langle0|T\phi_\alpha(x_1)\phi_{\alpha'}(x_2)|V\rangle=\frac{\varepsilon_{\alpha\alpha'\tau\tau'}}{\sqrt{6}M}\chi_{_\mathcal{V}}(x_1,x_2)\varepsilon^\tau P^{\tau'}
,
\end{eqnarray}
where $\varepsilon$ is the polarization vector of $1^+$.

The corresponding B-S equation should be
\begin{eqnarray} \label{4-dim-BS42}
\varepsilon_{\beta \beta' \omega\sigma}\chi_{_\mathcal{V}}(p)\varepsilon^\sigma P^\omega
=\Delta_{1\mu\beta}\int{d^4{q}\over(2\pi)^4}\,K^{\mu\mu'\nu\nu'}(P,p,q)
\varepsilon_{\nu\nu'\omega'\sigma'}\chi_{_\mathcal{V}}(q)\varepsilon^{\sigma'}P^{\omega'}\Delta_{2\mu'\beta'}\,,
\end{eqnarray}
and its form in three-momentum space
\begin{eqnarray} \label{3-dim-BS12}
\psi_{{}_\mathcal{V}}^{}({\bf |p|}) =&&\frac{
1}{12E^2E_1E_2[E^2-(E_1+E_2)^2](E+E_1-E_2)}\int{d^3\mathbf{p}'\over(2\pi)^3}\,
\psi_{{}_\mathcal{V}}^{}({\bf
|p|}')\nonumber\\&&\{E_2(E-E_1-E_2)[C_{\mathcal{V}1}|_{_{p_0=-E/2-E_1}}K_{v}(\mathbf{q})|F_V(\mathbf{q}^2)|^2+C_{\mathcal{V}2}|_{_{p_0=-E/2-E_1}}K_{s}(\mathbf{q})\kappa]\nonumber\\&&
+E_1(E+E_1+E_2)[C_{\mathcal{V}1}|_{_{p_0=E/2-E_2}}K_{v}(\mathbf{q})|F_V(\mathbf{q}^2)|^2+C_{\mathcal{V}2}|_{_{p_0=E/2-E_2}}K_{s}(\mathbf{q})\kappa]\}
.
\end{eqnarray}

The detailed expressions of $C_{\mathcal{V}1}$ and $C_{\mathcal{V}2}$ are also collected in Appendix.
\subsection{The B-S equation of $2^+$ state which is composed of two vectors}

 The B-S wave-function of $2^+$ state composed of two axial-vectors can be written as
 \begin{eqnarray} \label{4-dim-BS23}
\langle0|T\phi^\alpha(x_1)\phi^{\alpha'}(x_2)|V\rangle=\frac{1}{\sqrt{5}}\chi_{_\mathcal{T}}(x_1,x_2)\varepsilon^{\alpha\alpha'}
,
\end{eqnarray}
where $\varepsilon$ is the polarization vector of $2^+$.

The B-S equation can be expressed as
\begin{eqnarray} \label{4-dim-BS43}
\chi_{_\mathcal{T}}(p)\varepsilon_{\beta\beta'}
=\Delta_{1\mu\beta}\int{d^4{q}\over(2\pi)^4}\,K^{\mu\mu'\nu\nu'}(P,p,q)
\varepsilon_{\nu\nu'}\chi_{_\mathcal{T}}(q)\Delta_{2\mu'\beta'}\,.
\end{eqnarray}

Similarly one can obtain its three-momentum form
\begin{eqnarray} \label{3-dim-BS13}
\psi_{{}_\mathcal{T}}^{}({\bf |p|}) =&&\frac{
1}{10E_1E_2[E^2-(E_1+E_2)^2](E+E_1-E_2)}\int{d^3\mathbf{p}'\over(2\pi)^3}\,
\psi_{{}_\mathcal{T}}^{}({\bf
|p|}')\nonumber\\&&\{E_2(E-E_1-E_2)[C_{\mathcal{T}1}|_{_{p_0=-E/2-E_1}}K_{v}(\mathbf{q})|F_V(\mathbf{q}^2)|^2+C_{\mathcal{T}2}|_{_{p_0=-E/2-E_1}}K_{s}(\mathbf{q})\kappa]\nonumber\\&&
+E_1(E+E_1+E_2)[C_{\mathcal{T}1}|_{_{p_0=E/2-E_2}}K_{v}(\mathbf{q})|F_V(\mathbf{q}^2)|^2+C_{\mathcal{T}2}|_{_{p_0=E/2-E_2}}K_{s}(\mathbf{q})\kappa]\}
.
\end{eqnarray}

The expressions of $C_{\mathcal{T}1}$ and $C_{\mathcal{T}2}$ can be found  in Appendix.

\section{numerical results}
The B-S equations in Eqs.
(\ref{3-dim-BS11},\,\ref{3-dim-BS12},\,\ref{3-dim-BS13}) have
integral forms. Generally the standard way to solve an integral
equation is to discretize and perform algebraic operations.
Concretely,  we let $\bf |p|$ and $\bf |p'|$ take $n$ ( $n$ is
sufficient large) discrete values $Q_1$, $Q_2$,...$Q_n$ which
distribute with equal gap, then the integral equation is
transformed into $n$ coupled algebraic equations.
$\psi_{{}_\mathcal{N}}^{}(Q_1),\psi_{{}_\mathcal{N}}^{}(Q_2),...\psi_{{}_\mathcal{N}}^{}(Q_n)$
( the subscript $\mathcal{N}$ denotes $\mathcal{S}$, $\mathcal{V}$
or $\mathcal{T}$) constituents a column matrix  so these algebraic
equations can be regards as a matrix equation. We have explained
how to solve the equation in detail in our earlier
paper\cite{Ke:2019bkf}. It is noted that here we must deal with
the factor $E^2-(E_1+E_2)^2$ in the denominator to avoid the
singularity when $E>m_1+m_2$. {The following approach has been adopted : first,
$E^2-(E_1+E_2)^2$ is timed on both side in Eq. (\ref{3-dim-BS11}), (\ref{3-dim-BS12}) and (\ref{3-dim-BS13}),
and then $-(E_1+E_2)^2\psi_{_{\mathcal{N}}}(\bf |p|)$ ( the
subscript $\mathcal{N}$ denotes $\mathcal{S}$, $\mathcal{V}$ or
$\mathcal{T}$) is moved to right side of every equation, at last both sides of
every equation are divided by $E^2$.
}

To solve the B-S equation numerically some input parameters are
needed. At first we need to determine the masses of diquarks $cc$
and $bb$. In Refs.\cite{Li:2019ekr,Weng:2010rb,Yu:2006ty} the
theoretical values of heavy diquarks are presented. In
\cite{Li:2019ekr} and \cite{Weng:2010rb} the authors fixed the
mass of diquark $cc$ from the mass of $\Xi_{cc}$ so the predicted
diquark masses in \cite{Li:2019ekr} and \cite{Weng:2010rb} are
very close. In this work we will use the values in
\cite{Weng:2010rb} to do the calculations. The parameters $a=0.06$
GeV$^2$ and $\lambda=0.21$ GeV$^2$ are fixed in
Ref.\cite{Chang:2010kj}. In principle the strong coupling constant
$\alpha_s$ can be estimated using the expression
$\alpha_s(Q^2)=\frac{12\pi}{33-2N_f}\frac{1}{{\rm
ln}(\frac{Q^2}{\Lambda_{QCD}})}$ with $Q^2=m_D^2$ ($m_D$ is the
mass of diquark ) and $\Lambda_{QCD}=0.27$ GeV\cite{Chang:2010kj}.
We also need input value for the zero energy $V_0$. In
Refs.\cite{Wang:2009er,Wang:2007nb} for heavy quarkonia  the value
is about -0.4$\sim$-0.6 GeV so we will employ -0.5 GeV in our
calculation. We know very little about the parameter $\kappa$. In
Ref.\cite{Guo:1998ef,Guo:1996jj} the similar parameter for scalar
potential has the dimension of mass and is proportional to
$\Lambda_{QCD}$. Here we set it to be
$\Lambda_{QCD}\times\Lambda_{QCD}$, $\Lambda_{QCD}\times E$ or
$E\times E$ in our calculations ($E$ is the spectrum of the bound
state).

\begin{table}[!h]
\caption{The masses of diquarks in different references.}\label{Tab:t1}
\begin{ruledtabular}
\begin{tabular}{ccccc}
  &\cite{Li:2019ekr}
& \cite{Weng:2010rb}& \cite{Yu:2006ty}
\\\hline $m_{cc}$ (GeV) &3.303 &3.23& 3.52\\\hline $m_{bb}$ (GeV)
&9.830 &9.80& 10.28
\end{tabular}
\end{ruledtabular}
\end{table}

\begin{figure}
\begin{center}
\begin{tabular}{ccc}
\scalebox{0.53}{\includegraphics{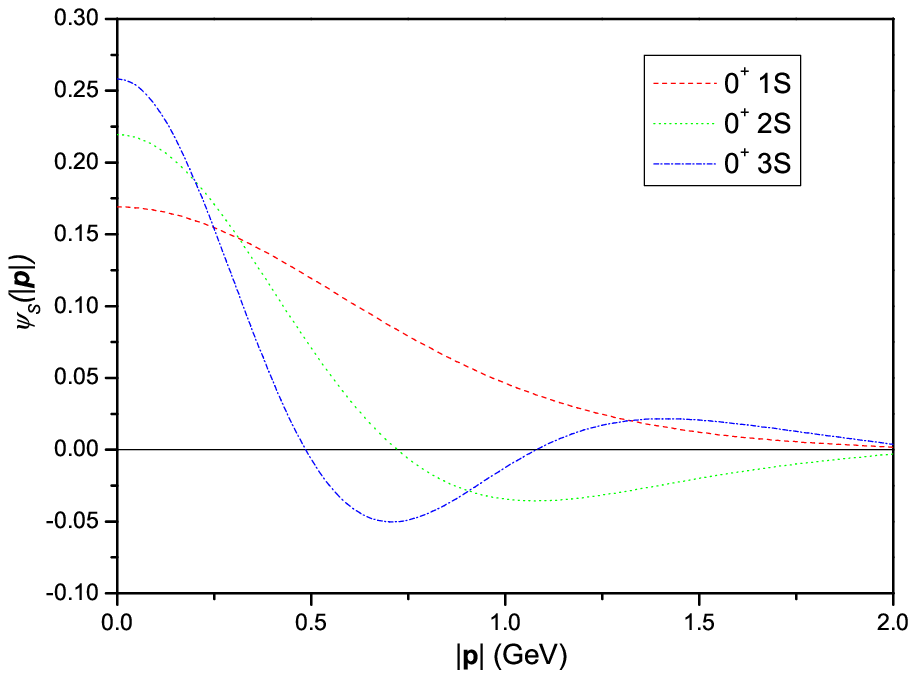}}\scalebox{0.53}{\includegraphics{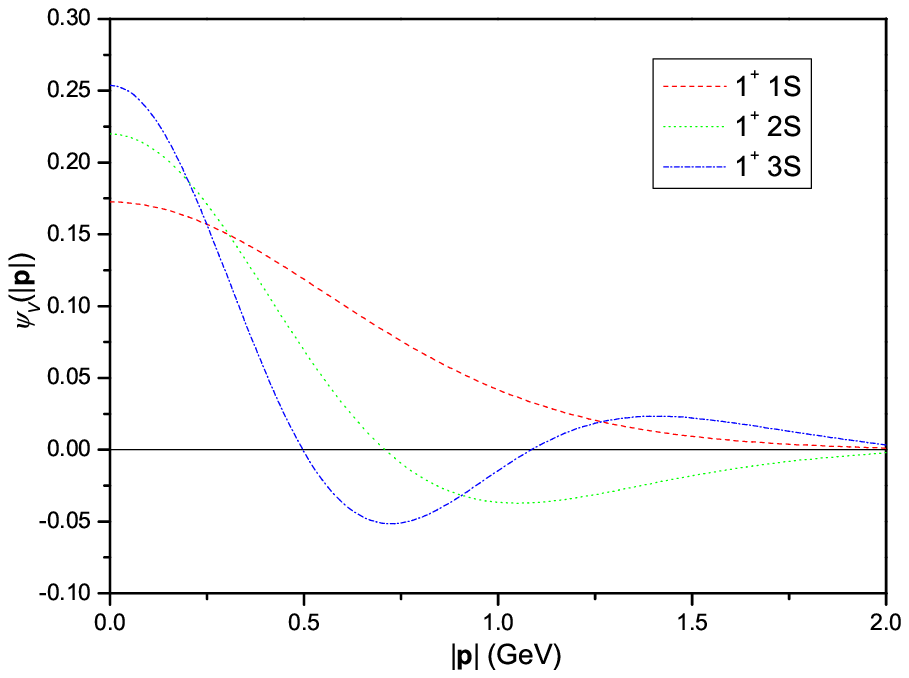}}\scalebox{0.53}{\includegraphics{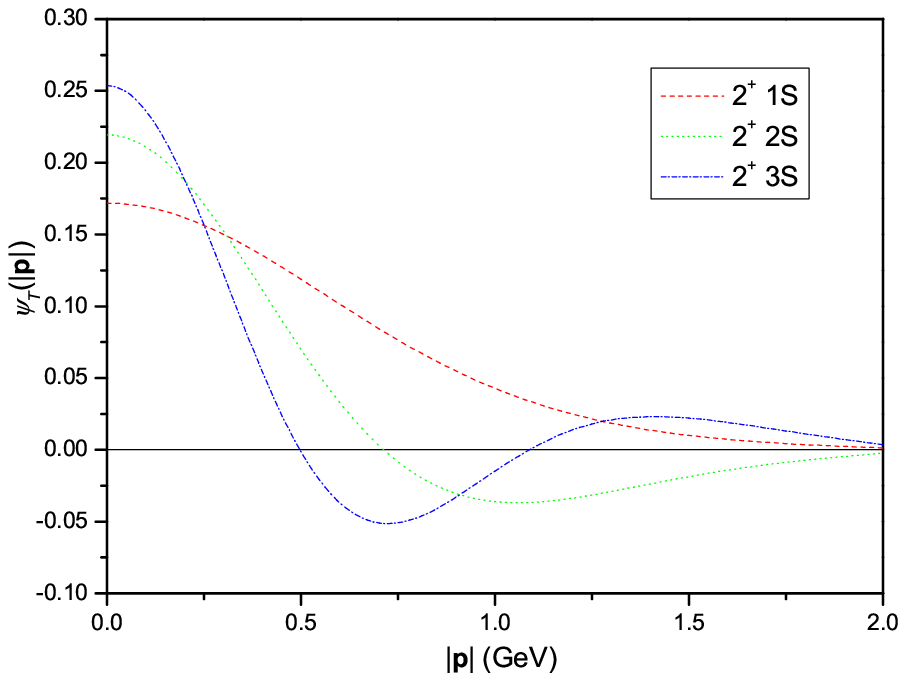}}
\end{tabular}
\end{center}
\caption{the wave functions of the $cc$-$\bar c\bar
c$ bound states}\label{wave}
\end{figure}

The spectra of the $cc$-$\bar c\bar c$ states we obtained are
present in table \ref{Tab04}. One can find the spectra of $1^+$
and $2^+$ are degenerate for different parameters. The mass of the second radial
excited state of $0^+$ is close to mass of $X(6900)$ when the
parameter $\kappa$ takes the value $E\times E$. In fact LHCb also
observed a board structure, using the Fig. 3(b) in Ref.\cite{Aaij2020}, the authors
\cite{Giron:2020wpx} estimated that its mass is $6490\pm 15$ GeV, whcih is close to that
of the first radial excited
state $0^+$. When one change the value of $\kappa$ the radial
excited state of $1^+$ or $2$ also can be consistent with data.
In Fig. \ref{wave} the wave functions of the $cc$-$\bar c\bar c$ bound states are depicted. One can find that the wave shapes of state
 with different quantum numbers are almost the same. It is noted the wave functions in Fig. \ref{wave} are not normalized.
Proper normalization is needed when one want to use them to
calculate the decay rates.

In table \ref{Tab04p} we list some theoretical
predictions\cite{Zhao:2020cfi,Chen:2020xwe,Wang:2020ols,Debastiani:2017msn,Bedolla:2019zwg}
on $cc$-$\bar c\bar c$ states. One can notice that the numerical
results covers a board range i.e., there is no conclusive answers
about the
spectra of $0^+$, $1^+$ and $2^+$. Certainly the changes of some input
parameters can alter the numerical results markedly. For
example if $m_{cc}=3.52$ GeV is used in our calculation all
numerical results will increase about 550 MeV. We also find that
the spectra of $1^+$ and $2^+$ are degenerate but there is a gap
with $0^+$ in Ref.\cite{Chen:2020xwe}, which is consistent with
ours. Nevertheless the mass splitting between  $0^+$ and  $1^+$ (
$2^+$) in our calculation is larger than those in
Ref.\cite{Chen:2020xwe}.

The spectra of  the $bb$-$\bar b\bar b$ bound states  was also
explored in Ref.\cite{Bedolla:2019zwg}. We also present our
predictions in table \ref{Tab05}. The wave functions are similar
to those in Fig. \ref{wave} so we ignore them here. Future
measurement of  $bb$-$\bar b\bar b$ state  will be crucial  for us
to determine the exact effective interaction between diquark and
antidiquark.

\begin{table}[!h]
\centering \caption{the spectra of the $cc$-$\bar c\bar c$ states
with $m_{cc}=3.23$ GeV\cite{Weng:2010rb} .} \label{Tab04}
\begin{ruledtabular}
\begin{tabular}{c|ccc|ccc|ccc}
\multirow{2}{*}{} & \multicolumn{3}{c|}{$0^+$} & \multicolumn{3}{c|}{$1^+$} & \multicolumn{3}{c}{$2^+$} \\
&  1S & 2S &   3S
&  1S & 2S &   3S
&  1S & 2S &   3S\\
\hline
$\Lambda_{QCD}\times\Lambda_{QCD}$ &6.270                          & 6.393                    & 6.441                   & 6.424           & 6.458           & 6.464          &6.424          & 6.458           &6.464         \\
$E\times\Lambda_{QCD}$ &6.271                          & 6.411                   & 6.477                   & 6.435           & 6.502           & 6.536          & 6.435           & 6.502           &6.536          \\
$E\times E$ &6.201                          & 6.575                   & 6.897                   & 6.396           & 6.799           & 7.148          &6.391          & 6.794           & 7.148          \\
\end{tabular}
\end{ruledtabular}
\end{table}

\begin{table}[!h]
\centering
\caption{the spectra of the $cc$-$\bar c\bar c$ states in different references
.}
\label{Tab04p}
\begin{ruledtabular}
\begin{tabular}{c|ccc|ccc|ccc}
\multirow{2}{*}{} & \multicolumn{3}{c|}{$0^+$} & \multicolumn{3}{c|}{$1^+$} & \multicolumn{3}{c}{$2^+$} \\
&  1S & 2S &   3S
&  1S & 2S &   3S
&  1S & 2S &   3S\\
\hline
\cite{Chen:2020xwe} &6.44                          & -                    & -                   & 6.51           & -          & -          & 6.51           & -           & -      \\
\cite{Zhao:2020cfi} &6.480                         & 6.906                    & 7.250                  & 6.508          & 6.934           & 6.991          &6.565          & 6.991           &7.355         \\
\cite{Wang:2020ols} &5.99                          & 6.48                    & 6.94                   & 6.05           & 6.52          & 6.96          & 6.09           & 6.56           & 7.00      \\
\cite{Bedolla:2019zwg} &5.883                          & 6.573                    & 6.948                   & 6.120           & 6.669          & 7.016          & 6.246           & 6.739           & 7.071     \\
\cite{Debastiani:2017msn} &  5.9694                        &6.6633                    & -                   & 6.0209           &  6.6745          & -          &  6.1154        & 6.6981           &  -         \\
\end{tabular}
\end{ruledtabular}
\end{table}

\begin{table}[!h]
\centering \caption{the spectra of the $bb$-$\bar b\bar b$ states
with $m_{bb}=9.8$ GeV .} \label{Tab05}
\begin{ruledtabular}
\begin{tabular}{c|ccc|ccc|ccc}
\multirow{2}{*}{} & \multicolumn{3}{c|}{$0^+$} & \multicolumn{3}{c|}{$1^+$} & \multicolumn{3}{c}{$2^+$} \\
&  1S & 2S &   3S
&  1S & 2S &   3S
&  1S & 2S &   3S\\
\hline
$\Lambda_{QCD}\times\Lambda_{QCD}$ &19.429                          & 19.512                    & 19.557                  &19.557           &19.587         &19.597          & 19.557         &19.587          & 19.597       \\
$E\times\Lambda_{QCD}$  &19.428           &19.515         &19.565                   & 19.558           & 19.597           & 19.615         & 19.558          & 19.597           & 19.615          \\
$E\times E$  & 19.302          &19.591          & 19.845                    &19.409           & 19.728           &20.016          & 19.409           & 19.728           &20.016          \\
\end{tabular}
\end{ruledtabular}
\end{table}

\section{summary and discussion}

In this paper we mainly explore the possible $cc$-$\bar c\bar c$
bound states within Bethe-Salpeter framework and we compare our
theoretical result to the measurement of  $X(6900)$ when the
parameters are properly chosen. Since the one-gluon exchange
potential is attractive for color anti-triplet (triplet) state we
only consider the ground axial-vector diquark (antidiquark) of
color anti-triplet (triplet). The $J^P$ of the diquark-antidiquark
state is $0^+$, $1^+$ or $2^+$. In order to deduce the B-S
equation of diquark-antidiquark we need to know the interaction
(kernel) between them. Since the coupling between diquark and
gluon is not presented in  fundamental QCD Lagrangian, in general
it is different from that between quark and gluon, so the
interaction (kernel) between quark and antiquark cannot be
directly applied to diquark-antidiquark system. Instead, we employ
the effective interaction to derive the tree level one-gluon
exchange vortex for the diquark-antidiquark system. Then we modify
the effective interaction kernel for meson  to study the possible
diquark-antidiquark bound state. Adopting the kernel, we deduce
the B-S equations for $0^+$, $1^+$ and $2^+$ diquark-antidiquark
system. Using the parameters based on previous studies of meson
and baryon, we solve these equations
 in 3-dimensional momentum space. In our calculation, the parameter $\kappa$ we introduced in kernel is undetermined so we take different values in the calculation.

The spectra of $1^+$
and $2^+$ are degenerate for different parameters in our results. The mass of the second radial
excited state $0^+$ is close to mass of $X(6900)$ when the
parameter $\kappa$ takes the value $E\times E$. In fact LHCb also
observed a board structure whose mass is $6490\pm 15$ GeV,  whcih is close to the first radial excited state $0^+$.

Theoretical results on $cc$-$\bar c\bar c$
bound states\cite{Zhao:2020cfi,Wang:2020ols,Bedolla:2019zwg} indicate $X(6900)$ maybe is the radial excited of $0^+$, $1^+$ or $2^+$ $cc$-$\bar c\bar c$ state.
However, LHCb only reported two structures in the mass region of $J/\psi$ pair between 6.2 to 7.4 GeV, which are less than the theoretical prediction for the number of possible states in $cc$-$\bar c\bar c$ system. In principle, If $X(6900)$ really is a $cc$-$\bar c\bar c$ bound state, other $cc$-$\bar c\bar c$ states predicted in theory should be seen in the experiment except that most states are degenerate.
Due to the lack of experimental evidence of a definite diqaurk and
antidiquark bound state, we are not able to completely fix free
parameters in our model, which causes some uncertainties in the
theoretical predictions. To overcome this issue, we will suggest
our experimental colleagues to measure the structures in the mass
region of $J/\psi$ pair between 6.2 to 7.4 GeV in depth. Using
present parameters we also calculate the spectra of the $bb$-$\bar
b\bar b$ bound states. In the future, more accurate data will help
us to elucidate the exact form of effective interaction between
diquark and antidiquark, and deepen our understanding of exotic
bound state systems.

\section*{Acknowledgement}
This work is supported by the National Natural Science Foundation
of China (NNSFC) under the contract No. 12075167 and 11975165. We
would like to
 thank Prof. Xue-Qian Li and Prof. Guo-Li Wang for their
 suggestions and useful discussions.

\appendix
\section{some detailed expressions}
\begin{eqnarray} C_{\mathcal{S}1}=&&\frac{7 {k_v}^2 {E}^6}{16 {m_1}^2 {m_2}^2}+\frac{3 {k_v} {E}^6}{16 {m_1}^2 {m_2}^2}+\frac{3 {E}^6}{64 {m_1}^2 {m_2}^2}-\frac{21
   {k_v}^2 {p_0}^2 {E}^4}{4 {m_1}^2 {m_2}^2}-\frac{9 {k_v} {p_0}^2 {E}^4}{4 {m_1}^2 {m_2}^2}-\frac{9 {p_0}^2 {E}^4}{16 {m_1}^2
   {m_2}^2}\nonumber\\&&-\frac{11 {k_v}^2 {\mathbf{p}\cdot\mathbf{q}} {E}^4}{8 {m_1}^2 {m_2}^2}-\frac{15 {k_v} {\mathbf{p}\cdot\mathbf{q}} {E}^4}{8 {m_1}^2
   {m_2}^2}-\frac{{\mathbf{p}\cdot\mathbf{q}} {E}^4}{4 {m_1}^2 {m_2}^2}+\frac{9 {k_v}^2 {\mathbf{p}^2} {E}^4}{16 {m_1}^2 {m_2}^2}+\frac{{k_v} {\mathbf{p}^2}
   {E}^4}{8 {m_1}^2 {m_2}^2}+\frac{{\mathbf{p}^2} {E}^4}{4 {m_1}^2 {m_2}^2}\nonumber\\&&+\frac{{k_v}^2 {\mathbf{q}^2} {E}^4}{16 {m_1}^2
   {m_2}^2}+\frac{{\mathbf{q}^2} {E}^4}{16 {m_1}^2 {m_2}^2}-\frac{7 {k_v}^2 {E}^4}{4 {m_1}^2}-\frac{3 {k_v} {E}^4}{4 {m_1}^2}-\frac{3
   {E}^4}{16 {m_1}^2}-\frac{7 {k_v}^2 {E}^4}{4 {m_2}^2}-\frac{3 {k_v} {E}^4}{4 {m_2}^2}\nonumber\\&&-\frac{3 {E}^4}{16 {m_2}^2}-\frac{7 {k_v}^2
   {p_0} {E}^3}{{m_1}^2}-\frac{3 {k_v} {p_0} {E}^3}{{m_1}^2}-\frac{3 {p_0} {E}^3}{4 {m_1}^2}+\frac{7 {k_v}^2 {p_0}
   {E}^3}{{m_2}^2}+\frac{3 {k_v} {p_0} {E}^3}{{m_2}^2}+\frac{3 {p_0} {E}^3}{4 {m_2}^2}\nonumber\\&&+\frac{21 {k_v}^2 {p_0}^4 {E}^2}{{m_1}^2
   {m_2}^2}+\frac{9 {k_v} {p_0}^4 {E}^2}{{m_1}^2 {m_2}^2}+\frac{9 {p_0}^4 {E}^2}{4 {m_1}^2 {m_2}^2}+10 {k_v}^2 {E}^2+\frac{3
   {k_v}^2 {\mathbf{p}\cdot\mathbf{q}}^2 {E}^2}{2 {m_1}^2 {m_2}^2}+\frac{3 {k_v} {\mathbf{p}\cdot\mathbf{q}}^2 {E}^2}{2 {m_1}^2 {m_2}^2}\nonumber\\&&+\frac{{\mathbf{p}\cdot\mathbf{q}}^2
   {E}^2}{2 {m_1}^2 {m_2}^2}-\frac{3 {k_v}^2 \mathbf{p}^4 {E}^2}{{m_1}^2 {m_2}^2}-\frac{5 {k_v} \mathbf{p}^4 {E}^2}{2 {m_1}^2
   {m_2}^2}+\frac{\mathbf{p}^4 {E}^2}{4 {m_1}^2 {m_2}^2}+6 {k_v} {E}^2-\frac{{k_v}^2 {p_0}^2 {\mathbf{p}\cdot\mathbf{q}} {E}^2}{{m_1}^2
   {m_2}^2}\nonumber\\&&+\frac{3 {k_v} {p_0}^2 {\mathbf{p}\cdot\mathbf{q}} {E}^2}{{m_1}^2 {m_2}^2}-\frac{{k_v}^2 {\mathbf{p}\cdot\mathbf{q}} {E}^2}{2 {m_1}^2}-\frac{{k_v}
   {\mathbf{p}\cdot\mathbf{q}} {E}^2}{2 {m_1}^2}-\frac{{\mathbf{p}\cdot\mathbf{q}} {E}^2}{2 {m_1}^2}-\frac{{k_v}^2 {\mathbf{p}\cdot\mathbf{q}} {E}^2}{2 {m_2}^2}-\frac{{k_v}
   {\mathbf{p}\cdot\mathbf{q}} {E}^2}{2 {m_2}^2}\nonumber\\&&-\frac{{\mathbf{p}\cdot\mathbf{q}} {E}^2}{2 {m_2}^2}-\frac{33 {k_v}^2 {p_0}^2 {\mathbf{p}^2} {E}^2}{2 {m_1}^2
   {m_2}^2}-\frac{5 {k_v} {p_0}^2 {\mathbf{p}^2} {E}^2}{{m_1}^2 {m_2}^2}-\frac{2 {p_0}^2 {\mathbf{p}^2} {E}^2}{{m_1}^2 {m_2}^2}-\frac{2 {k_v}^2
   {\mathbf{p}\cdot\mathbf{q}} {\mathbf{p}^2} {E}^2}{{m_1}^2 {m_2}^2}\nonumber\\&&-\frac{6 {k_v} {\mathbf{p}\cdot\mathbf{q}} {\mathbf{p}^2} {E}^2}{{m_1}^2 {m_2}^2}-\frac{{\mathbf{p}\cdot\mathbf{q}}
   {\mathbf{p}^2} {E}^2}{{m_1}^2 {m_2}^2}+\frac{3 {k_v}^2 {\mathbf{p}^2} {E}^2}{4 {m_1}^2}+\frac{{k_v} {\mathbf{p}^2} {E}^2}{2 {m_1}^2}+\frac{3
   {\mathbf{p}^2} {E}^2}{4 {m_1}^2}+\frac{3 {k_v}^2 {\mathbf{p}^2} {E}^2}{4 {m_2}^2}+\frac{{k_v} {\mathbf{p}^2} {E}^2}{2 {m_2}^2}\nonumber\\&&+\frac{3 {\mathbf{p}^2}
   {E}^2}{4 {m_2}^2}-\frac{{k_v}^2 {p_0}^2 {\mathbf{q}^2} {E}^2}{2 {m_1}^2 {m_2}^2}-\frac{{p_0}^2 {\mathbf{q}^2} {E}^2}{2 {m_1}^2
   {m_2}^2}+\frac{{k_v}^2 {\mathbf{p}^2} {\mathbf{q}^2} {E}^2}{2 {m_1}^2 {m_2}^2}+\frac{{\mathbf{p}^2} {\mathbf{q}^2} {E}^2}{2 {m_1}^2 {m_2}^2}-\frac{{k_v}^2
   {\mathbf{q}^2} {E}^2}{4 {m_1}^2}-\frac{{\mathbf{q}^2} {E}^2}{4 {m_1}^2}\nonumber\\&&-\frac{{k_v}^2 {\mathbf{q}^2} {E}^2}{4 {m_2}^2}-\frac{{\mathbf{q}^2} {E}^2}{4
   {m_2}^2}+\frac{9 {E}^2}{2}+\frac{28 {k_v}^2 {p_0}^3 {E}}{{m_1}^2}+\frac{12 {k_v} {p_0}^3 {E}}{{m_1}^2}+\frac{3 {p_0}^3
   {E}}{{m_1}^2}-\frac{28 {k_v}^2 {p_0}^3 {E}}{{m_2}^2}\nonumber\\&&-\frac{12 {k_v} {p_0}^3 {E}}{{m_2}^2}-\frac{3 {p_0}^3
   {E}}{{m_2}^2}-\frac{14 {k_v}^2 {p_0} {\mathbf{p}\cdot\mathbf{q}} {E}}{{m_1}^2}-\frac{10 {k_v} {p_0} {\mathbf{p}\cdot\mathbf{q}} {E}}{{m_1}^2}-\frac{3 {p_0}
   {\mathbf{p}\cdot\mathbf{q}} {E}}{{m_1}^2}+\frac{14 {k_v}^2 {p_0} {\mathbf{p}\cdot\mathbf{q}} {E}}{{m_2}^2}\nonumber\\&&+\frac{10 {k_v} {p_0} {\mathbf{p}\cdot\mathbf{q}}
   {E}}{{m_2}^2}+\frac{3 {p_0} {\mathbf{p}\cdot\mathbf{q}} {E}}{{m_2}^2}-\frac{13 {k_v}^2 {p_0} {\mathbf{p}^2} {E}}{{m_1}^2}-\frac{2 {k_v} {p_0}
   {\mathbf{p}^2} {E}}{{m_1}^2}+\frac{{p_0} {\mathbf{p}^2} {E}}{{m_1}^2}+\frac{13 {k_v}^2 {p_0} {\mathbf{p}^2} {E}}{{m_2}^2}\nonumber\\&&+\frac{2 {k_v} {p_0}
   {\mathbf{p}^2} {E}}{{m_2}^2}-\frac{{p_0} {\mathbf{p}^2} {E}}{{m_2}^2}-\frac{{k_v}^2 {p_0} {\mathbf{q}^2} {E}}{{m_1}^2}-\frac{{p_0} {\mathbf{q}^2}
   {E}}{{m_1}^2}+\frac{{k_v}^2 {p_0} {\mathbf{q}^2} {E}}{{m_2}^2}+\frac{{p_0} {\mathbf{q}^2} {E}}{{m_2}^2}-\frac{28 {k_v}^2
   {p_0}^6}{{m_1}^2 {m_2}^2}\nonumber\\&&-\frac{12 {k_v} {p_0}^6}{{m_1}^2 {m_2}^2}-\frac{3 {p_0}^6}{{m_1}^2 {m_2}^2}+\frac{28 {k_v}^2
   {p_0}^4}{{m_1}^2}+\frac{12 {k_v} {p_0}^4}{{m_1}^2}+\frac{3 {p_0}^4}{{m_1}^2}+\frac{28 {k_v}^2 {p_0}^4}{{m_2}^2}+\frac{12 {k_v}
   {p_0}^4}{{m_2}^2}+\frac{3 {p_0}^4}{{m_2}^2}\nonumber\\&&+\frac{7 {k_v}^2 \mathbf{p}^6}{{m_1}^2 {m_2}^2}-40 {k_v}^2 {p_0}^2-24 {k_v} {p_0}^2-18
   {p_0}^2-\frac{6 {k_v}^2 {p_0}^2 {\mathbf{p}\cdot\mathbf{q}}^2}{{m_1}^2 {m_2}^2}-\frac{6 {k_v} {p_0}^2 {\mathbf{p}\cdot\mathbf{q}}^2}{{m_1}^2 {m_2}^2}-\frac{2 {p_0}^2
   {\mathbf{p}\cdot\mathbf{q}}^2}{{m_1}^2 {m_2}^2}\nonumber\\&&+\frac{6 {k_v}^2 {\mathbf{p}\cdot\mathbf{q}}^2}{{m_1}^2}+\frac{6 {k_v} {\mathbf{p}\cdot\mathbf{q}}^2}{{m_1}^2}+\frac{2
   {\mathbf{p}\cdot\mathbf{q}}^2}{{m_1}^2}+\frac{6 {k_v}^2 {\mathbf{p}\cdot\mathbf{q}}^2}{{m_2}^2}+\frac{6 {k_v} {\mathbf{p}\cdot\mathbf{q}}^2}{{m_2}^2}+\frac{2 {\mathbf{p}\cdot\mathbf{q}}^2}{{m_2}^2}-\frac{36
   {k_v}^2 {p_0}^2 \mathbf{p}^4}{{m_1}^2 {m_2}^2}\nonumber\\&&-\frac{6 {k_v} {p_0}^2 \mathbf{p}^4}{{m_1}^2 {m_2}^2}-\frac{{p_0}^2 \mathbf{p}^4}{{m_1}^2
   {m_2}^2}+\frac{14 {k_v}^2 {\mathbf{p}\cdot\mathbf{q}} \mathbf{p}^4}{{m_1}^2 {m_2}^2}+\frac{6 {k_v} {\mathbf{p}\cdot\mathbf{q}} \mathbf{p}^4}{{m_1}^2 {m_2}^2}+\frac{7 {k_v}^2
   \mathbf{p}^4}{{m_1}^2}+\frac{7 {k_v}^2 \mathbf{p}^4}{{m_2}^2}+\frac{26 {k_v}^2 {p_0}^4 {\mathbf{p}\cdot\mathbf{q}}}{{m_1}^2 {m_2}^2}\nonumber\\&&+\frac{18 {k_v} {p_0}^4
   {\mathbf{p}\cdot\mathbf{q}}}{{m_1}^2 {m_2}^2}+\frac{4 {p_0}^4 {\mathbf{p}\cdot\mathbf{q}}}{{m_1}^2 {m_2}^2}+20 {k_v}^2 {\mathbf{p}\cdot\mathbf{q}}-\frac{26 {k_v}^2 {p_0}^2
   {\mathbf{p}\cdot\mathbf{q}}}{{m_1}^2}-\frac{18 {k_v} {p_0}^2 {\mathbf{p}\cdot\mathbf{q}}}{{m_1}^2}-\frac{4 {p_0}^2 {\mathbf{p}\cdot\mathbf{q}}}{{m_1}^2}\nonumber\\&&-\frac{26 {k_v}^2 {p_0}^2
   {\mathbf{p}\cdot\mathbf{q}}}{{m_2}^2}-\frac{18 {k_v} {p_0}^2 {\mathbf{p}\cdot\mathbf{q}}}{{m_2}^2}-\frac{4 {p_0}^2 {\mathbf{p}\cdot\mathbf{q}}}{{m_2}^2}+12 {k_v} {\mathbf{p}\cdot\mathbf{q}}+6
   {\mathbf{p}\cdot\mathbf{q}}+\frac{57 {k_v}^2 {p_0}^4 {\mathbf{p}^2}}{{m_1}^2 {m_2}^2}\nonumber\\&&+\frac{18 {k_v} {p_0}^4 {\mathbf{p}^2}}{{m_1}^2 {m_2}^2}+\frac{4 {p_0}^4
   {\mathbf{p}^2}}{{m_1}^2 {m_2}^2}+10 {k_v}^2 {\mathbf{p}^2}-\frac{29 {k_v}^2 {p_0}^2 {\mathbf{p}^2}}{{m_1}^2}-\frac{6 {k_v} {p_0}^2
   {\mathbf{p}^2}}{{m_1}^2}-\frac{{p_0}^2 {\mathbf{p}^2}}{{m_1}^2}-\frac{29 {k_v}^2 {p_0}^2 {\mathbf{p}^2}}{{m_2}^2}\nonumber\\&&-\frac{6 {k_v} {p_0}^2
   {\mathbf{p}^2}}{{m_2}^2}-\frac{{p_0}^2 {\mathbf{p}^2}}{{m_2}^2}+\frac{6 {k_v}^2 {\mathbf{p}\cdot\mathbf{q}}^2 {\mathbf{p}^2}}{{m_1}^2 {m_2}^2}+\frac{6 {k_v} {\mathbf{p}\cdot\mathbf{q}}^2
   {\mathbf{p}^2}}{{m_1}^2 {m_2}^2}+\frac{2 {\mathbf{p}\cdot\mathbf{q}}^2 {\mathbf{p}^2}}{{m_1}^2 {m_2}^2}+6 {k_v} {\mathbf{p}^2}-\frac{40 {k_v}^2 {p_0}^2 {\mathbf{p}\cdot\mathbf{q}}
   {\mathbf{p}^2}}{{m_1}^2 {m_2}^2}\nonumber\\&&-\frac{24 {k_v} {p_0}^2 {\mathbf{p}\cdot\mathbf{q}} {\mathbf{p}^2}}{{m_1}^2 {m_2}^2}-\frac{4 {p_0}^2 {\mathbf{p}\cdot\mathbf{q}} {\mathbf{p}^2}}{{m_1}^2
   {m_2}^2}+\frac{14 {k_v}^2 {\mathbf{p}\cdot\mathbf{q}} {\mathbf{p}^2}}{{m_1}^2}+\frac{6 {k_v} {\mathbf{p}\cdot\mathbf{q}} {\mathbf{p}^2}}{{m_1}^2}+\frac{14 {k_v}^2 {\mathbf{p}\cdot\mathbf{q}}
   {\mathbf{p}^2}}{{m_2}^2}\nonumber\\&&+\frac{6 {k_v} {\mathbf{p}\cdot\mathbf{q}} {\mathbf{p}^2}}{{m_2}^2}+6 {\mathbf{p}^2}+\frac{{k_v}^2 {p_0}^4 {\mathbf{q}^2}}{{m_1}^2 {m_2}^2}+\frac{{p_0}^4
   {\mathbf{q}^2}}{{m_1}^2 {m_2}^2}+10 {k_v}^2 {\mathbf{q}^2}-\frac{{k_v}^2 {p_0}^2 {\mathbf{q}^2}}{{m_1}^2}-\frac{{p_0}^2 {\mathbf{q}^2}}{{m_1}^2}-\frac{{k_v}^2
   {p_0}^2 {\mathbf{q}^2}}{{m_2}^2}\nonumber\\&&-\frac{{p_0}^2 {\mathbf{q}^2}}{{m_2}^2}+\frac{{k_v}^2 \mathbf{p}^4 {\mathbf{q}^2}}{{m_1}^2 {m_2}^2}+\frac{\mathbf{p}^4
   {\mathbf{q}^2}}{{m_1}^2 {m_2}^2}+6 {k_v} {\mathbf{q}^2}-\frac{2 {k_v}^2 {p_0}^2 {\mathbf{p}^2} {\mathbf{q}^2}}{{m_1}^2 {m_2}^2}-\frac{2 {p_0}^2 {\mathbf{p}^2}
   {\mathbf{q}^2}}{{m_1}^2 {m_2}^2}\nonumber\\&&+\frac{{k_v}^2 {\mathbf{p}^2} {\mathbf{q}^2}}{{m_1}^2}+\frac{{\mathbf{p}^2} {\mathbf{q}^2}}{{m_1}^2}+\frac{{k_v}^2 {\mathbf{p}^2}
   {\mathbf{q}^2}}{{m_2}^2}+\frac{{\mathbf{p}^2} {\mathbf{q}^2}}{{m_2}^2}+6 {\mathbf{q}^2}
\end{eqnarray}

 \begin{eqnarray}
 C_{\mathcal{S}2}=&&\frac{{E}^4}{16 {m_1}^2 {m_2}^2}-\frac{{E}^2 {p_0}^2}{2 {m_1}^2 {m_2}^2}+\frac{{E}^2 {\mathbf{p}^2}}{2 {m_1}^2 {m_2}^2}-\frac{{E}^2}{4
   {m_1}^2}-\frac{{E}^2}{4 {m_2}^2}-\frac{{E} {p_0}}{{m_1}^2}+\frac{{E} {p_0}}{{m_2}^2}+\frac{{p_0}^4}{{m_1}^2 {m_2}^2}\nonumber\\&&-\frac{2
   {p_0}^2 {\mathbf{p}^2}}{{m_1}^2 {m_2}^2}+\frac{{\mathbf{p}^4}}{{m_1}^2
   {m_2}^2}-\frac{{p_0}^2}{{m_1}^2}+\frac{{\mathbf{p}^2}}{{m_1}^2}-\frac{{p_0}^2}{{m_2}^2}+\frac{{\mathbf{p}^2}}{{m_2}^2}+4
 \end{eqnarray}

\begin{eqnarray}
 C_{\mathcal{V}1}=&&\frac{{E}^4 {k_v}^2 {\mathbf{p}\cdot\mathbf{q}}}{{m_1}^2}+\frac{{E}^4 {k_v}^2 {\mathbf{p}^2}}{{m_1}^2}+\frac{{E}^4 {k_v}^2
   {\mathbf{p}\cdot\mathbf{q}}}{{m_2}^2}+\frac{{E}^4 {k_v}^2 {\mathbf{p}^2}}{{m_2}^2}+\frac{3 {E}^4 {k_v} {\mathbf{p}\cdot\mathbf{q}}}{2 {m_1}^2}+\frac{{E}^4 {k_v}
   {\mathbf{p}^2}}{2 {m_1}^2}+\frac{3 {E}^4 {k_v} {\mathbf{p}\cdot\mathbf{q}}}{2 {m_2}^2}\nonumber\\&&+\frac{{E}^4 {k_v} {\mathbf{p}^2}}{2 {m_2}^2}+\frac{5 {E}^4 {\mathbf{p}^2}}{2
   {m_1}^2}+\frac{5 {E}^4 {\mathbf{p}^2}}{2 {m_2}^2}+6 {E}^4+\frac{4 {E}^3 {k_v}^2 {p_0} {\mathbf{p}\cdot\mathbf{q}}}{{m_1}^2}+\frac{4 {E}^3 {k_v}^2
   {p_0} {\mathbf{p}^2}}{{m_1}^2}-\frac{4 {E}^3 {k_v}^2 {p_0} {\mathbf{p}\cdot\mathbf{q}}}{{m_2}^2}\nonumber\\&&-\frac{4 {E}^3 {k_v}^2 {p_0}
   {\mathbf{p}^2}}{{m_2}^2}+\frac{10 {E}^3 {k_v} {p_0} {\mathbf{p}\cdot\mathbf{q}}}{{m_1}^2}+\frac{6 {E}^3 {k_v} {p_0} {\mathbf{p}^2}}{{m_1}^2}-\frac{10
   {E}^3 {k_v} {p_0} {\mathbf{p}\cdot\mathbf{q}}}{{m_2}^2}-\frac{6 {E}^3 {k_v} {p_0} {\mathbf{p}^2}}{{m_2}^2}\nonumber\\&&+\frac{4 {E}^3 {p_0}
   {\mathbf{p}\cdot\mathbf{q}}}{{m_1}^2}+\frac{2 {E}^3 {p_0} {\mathbf{p}^2}}{{m_1}^2}-\frac{4 {E}^3 {p_0} {\mathbf{p}\cdot\mathbf{q}}}{{m_2}^2}-\frac{2 {E}^3 {p_0}
   {\mathbf{p}^2}}{{m_2}^2}+\frac{4 {E}^2 {k_v}^2 {p_0}^2 {\mathbf{p}\cdot\mathbf{q}}}{{m_1}^2}+\frac{4 {E}^2 {k_v}^2 {p_0}^2
   {\mathbf{p}^2}}{{m_1}^2}\nonumber\\&&-\frac{{E}^2 {k_v}^2 {\mathbf{p}\cdot\mathbf{q}}^2}{{m_1}^2}-\frac{4 {E}^2 {k_v}^2 {\mathbf{p}\cdot\mathbf{q}} {\mathbf{p}^2}}{{m_1}^2}-\frac{2 {E}^2
   {k_v}^2 \mathbf{p}^4}{{m_1}^2}-\frac{{E}^2 {k_v}^2 {\mathbf{p}^2} {\mathbf{q}^2}}{{m_1}^2}+\frac{4 {E}^2 {k_v}^2 {p_0}^2
   {\mathbf{p}\cdot\mathbf{q}}}{{m_2}^2}\nonumber\\&&+\frac{4 {E}^2 {k_v}^2 {p_0}^2 {\mathbf{p}^2}}{{m_2}^2}-\frac{{E}^2 {k_v}^2 {\mathbf{p}\cdot\mathbf{q}}^2}{{m_2}^2}-\frac{4 {E}^2
   {k_v}^2 {\mathbf{p}\cdot\mathbf{q}} {\mathbf{p}^2}}{{m_2}^2}-\frac{2 {E}^2 {k_v}^2 \mathbf{p}^4}{{m_2}^2}-\frac{{E}^2 {k_v}^2 {\mathbf{p}^2} {\mathbf{q}^2}}{{m_2}^2}\nonumber\\&&-8
   {E}^2 {k_v}^2 {\mathbf{p}\cdot\mathbf{q}}-4 {E}^2 {k_v}^2 {\mathbf{p}^2}-4 {E}^2 {k_v}^2 {\mathbf{q}^2}+\frac{14 {E}^2 {k_v} {p_0}^2
   {\mathbf{p}\cdot\mathbf{q}}}{{m_1}^2}+\frac{10 {E}^2 {k_v} {p_0}^2 {\mathbf{p}^2}}{{m_1}^2}\nonumber\\&&-\frac{3 {E}^2 {k_v} {\mathbf{p}\cdot\mathbf{q}}^2}{{m_1}^2}-\frac{12 {E}^2
   {k_v} {\mathbf{p}\cdot\mathbf{q}} {\mathbf{p}^2}}{{m_1}^2}-\frac{6 {E}^2 {k_v} \mathbf{p}^4}{{m_1}^2}-\frac{3 {E}^2 {k_v} {\mathbf{p}^2}
   {\mathbf{q}^2}}{{m_1}^2}+\frac{14 {E}^2 {k_v} {p_0}^2 {\mathbf{p}\cdot\mathbf{q}}}{{m_2}^2}\nonumber\\&&+\frac{10 {E}^2 {k_v} {p_0}^2 {\mathbf{p}^2}}{{m_2}^2}-\frac{3
   {E}^2 {k_v} {\mathbf{p}\cdot\mathbf{q}}^2}{{m_2}^2}-\frac{12 {E}^2 {k_v} {\mathbf{p}\cdot\mathbf{q}} {\mathbf{p}^2}}{{m_2}^2}-\frac{6 {E}^2 {k_v}
   \mathbf{p}^4}{{m_2}^2}-\frac{3 {E}^2 {k_v} {\mathbf{p}^2} {\mathbf{q}^2}}{{m_2}^2}\nonumber\\&&-24 {E}^2 {k_v} {\mathbf{p}\cdot\mathbf{q}}-12 {E}^2 {k_v} {\mathbf{p}^2}-12
   {E}^2 {k_v} {\mathbf{q}^2}+\frac{8 {E}^2 {p_0}^2 {\mathbf{p}\cdot\mathbf{q}}}{{m_1}^2}-\frac{6 {E}^2 {p_0}^2 {\mathbf{p}^2}}{{m_1}^2}-\frac{{E}^2
   {\mathbf{p}\cdot\mathbf{q}}^2}{{m_1}^2}\nonumber\\&&-\frac{2 {E}^2 {\mathbf{p}\cdot\mathbf{q}} {\mathbf{p}^2}}{{m_1}^2}+\frac{{E}^2 {\mathbf{p}^2} {\mathbf{q}^2}}{{m_1}^2}+\frac{8 {E}^2 {p_0}^2
   {\mathbf{p}\cdot\mathbf{q}}}{{m_2}^2}-\frac{6 {E}^2 {p_0}^2 {\mathbf{p}^2}}{{m_2}^2}-\frac{{E}^2 {\mathbf{p}\cdot\mathbf{q}}^2}{{m_2}^2}-\frac{2 {E}^2 {\mathbf{p}\cdot\mathbf{q}}
   {\mathbf{p}^2}}{{m_2}^2}\nonumber\\&&+\frac{{E}^2 {\mathbf{p}^2} {\mathbf{q}^2}}{{m_2}^2}-24 {E}^2 {p_0}^2+2 {E}^2 {\mathbf{p}^2}+2 {E}^2 {\mathbf{q}^2}
\end{eqnarray}

 \begin{eqnarray}
 C_{\mathcal{V}2}=\frac{2 {E}^2 {\mathbf{p}^2}}{{m_1}^2}+\frac{2 {E}^2 {\mathbf{p}^2}}{{m_2}^2}+6 {E}^2
 \end{eqnarray}

\begin{eqnarray}
 C_{\mathcal{T}1}=&&\frac{2 {k_v}^2 \mathbf{p}^6}{{m_1}^2 {m_2}^2}-\frac{4 {k_v} \mathbf{p}^6}{3 {m_1}^2 {m_2}^2}-\frac{2 \mathbf{p}^6}{3 {m_1}^2 {m_2}^2}-\frac{10
   {k_v}^2 {p_0}^2 \mathbf{p}^4}{3 {m_1}^2 {m_2}^2}+\frac{4 {k_v} {p_0}^2 \mathbf{p}^4}{{m_1}^2 {m_2}^2}-\frac{4 {p_0}^2 \mathbf{p}^4}{3 {m_1}^2
   {m_2}^2}+\frac{4 {k_v}^2 {\mathbf{p}\cdot\mathbf{q}} \mathbf{p}^4}{{m_1}^2 {m_2}^2}\nonumber\\&&-\frac{4 {k_v} {\mathbf{p}\cdot\mathbf{q}} \mathbf{p}^4}{3 {m_1}^2 {m_2}^2}-\frac{{k_v}^2
   {\mathbf{q}^2} \mathbf{p}^4}{3 {m_1}^2 {m_2}^2}-\frac{2 {k_v} {\mathbf{q}^2} \mathbf{p}^4}{3 {m_1}^2 {m_2}^2}+\frac{{\mathbf{q}^2} \mathbf{p}^4}{3 {m_1}^2
   {m_2}^2}+\frac{3 {k_v}^2 \mathbf{p}^4}{{m_1}^2}-\frac{7 {k_v} \mathbf{p}^4}{3 {m_1}^2}-\frac{2 \mathbf{p}^4}{3 {m_1}^2}+\frac{3 {k_v}^2
   \mathbf{p}^4}{{m_2}^2}\nonumber\\&&-\frac{7 {k_v} \mathbf{p}^4}{3 {m_2}^2}+\frac{{E}^2 \mathbf{p}^4}{{m_1}^2 {m_2}^2}-\frac{{E}^2 {k_v}^2 \mathbf{p}^4}{2
   {m_1}^2 {m_2}^2}-\frac{{E}^2 {k_v} \mathbf{p}^4}{3 {m_1}^2 {m_2}^2}-\frac{2 \mathbf{p}^4}{3 {m_2}^2}+\frac{10 {k_v}^2 {\mathbf{p}^2}}{3}-\frac{10
   {k_v}^2 {p_0}^2 {\mathbf{p}^2}}{3 {m_1}^2}+\frac{5 {k_v} {p_0}^2 {\mathbf{p}^2}}{{m_1}^2}\nonumber\\&&-\frac{5 {p_0}^2 {\mathbf{p}^2}}{{m_1}^2}-\frac{10 {k_v}^2
   {p_0}^2 {\mathbf{p}^2}}{3 {m_2}^2}+\frac{5 {k_v} {p_0}^2 {\mathbf{p}^2}}{{m_2}^2}-\frac{5 {p_0}^2 {\mathbf{p}^2}}{{m_2}^2}+\frac{7 {k_v}^2 {\mathbf{p}\cdot\mathbf{q}}^2
   {\mathbf{p}^2}}{3 {m_1}^2 {m_2}^2}+\frac{2 {k_v} {\mathbf{p}\cdot\mathbf{q}}^2 {\mathbf{p}^2}}{3 {m_1}^2 {m_2}^2}-\frac{{\mathbf{p}\cdot\mathbf{q}}^2 {\mathbf{p}^2}}{3 {m_1}^2
   {m_2}^2}\nonumber\\&&-\frac{10 {k_v} {\mathbf{p}^2}}{3}-\frac{10 {E} {k_v}^2 {p_0} {\mathbf{p}^2}}{3 {m_1}^2}+\frac{5 {E} {p_0} {\mathbf{p}^2}}{3 {m_1}^2}+\frac{5
   {E} {k_v} {p_0} {\mathbf{p}^2}}{3 {m_1}^2}+\frac{10 {E} {k_v}^2 {p_0} {\mathbf{p}^2}}{3 {m_2}^2}-\frac{5 {E} {p_0} {\mathbf{p}^2}}{3
   {m_2}^2}-\frac{5 {E} {k_v} {p_0} {\mathbf{p}^2}}{3 {m_2}^2}\nonumber\\&&-\frac{4 {k_v}^2 {p_0}^2 {\mathbf{p}\cdot\mathbf{q}} {\mathbf{p}^2}}{{m_1}^2 {m_2}^2}+\frac{8
   {p_0}^2 {\mathbf{p}\cdot\mathbf{q}} {\mathbf{p}^2}}{3 {m_1}^2 {m_2}^2}+\frac{6 {k_v}^2 {\mathbf{p}\cdot\mathbf{q}} {\mathbf{p}^2}}{{m_1}^2}-\frac{10 {k_v} {\mathbf{p}\cdot\mathbf{q}} {\mathbf{p}^2}}{3
   {m_1}^2}+\frac{{\mathbf{p}\cdot\mathbf{q}} {\mathbf{p}^2}}{{m_1}^2}+\frac{6 {k_v}^2 {\mathbf{p}\cdot\mathbf{q}} {\mathbf{p}^2}}{{m_2}^2}\nonumber\\&&-\frac{10 {k_v} {\mathbf{p}\cdot\mathbf{q}} {\mathbf{p}^2}}{3
   {m_2}^2}-\frac{2 {E}^2 {\mathbf{p}\cdot\mathbf{q}} {\mathbf{p}^2}}{3 {m_1}^2 {m_2}^2}-\frac{{E}^2 {k_v}^2 {\mathbf{p}\cdot\mathbf{q}} {\mathbf{p}^2}}{3 {m_1}^2
   {m_2}^2}-\frac{2 {E}^2 {k_v} {\mathbf{p}\cdot\mathbf{q}} {\mathbf{p}^2}}{3 {m_1}^2 {m_2}^2}+\frac{{\mathbf{p}\cdot\mathbf{q}} {\mathbf{p}^2}}{{m_2}^2}+\frac{{k_v}^2 {p_0}^2
   {\mathbf{q}^2} {\mathbf{p}^2}}{3 {m_1}^2 {m_2}^2}\nonumber\\&&+\frac{2 {k_v} {p_0}^2 {\mathbf{q}^2} {\mathbf{p}^2}}{3 {m_1}^2 {m_2}^2}+\frac{{p_0}^2 {\mathbf{q}^2} {\mathbf{p}^2}}{3
   {m_1}^2 {m_2}^2}+\frac{{k_v}^2 {\mathbf{q}^2} {\mathbf{p}^2}}{6 {m_1}^2}-\frac{7 {k_v} {\mathbf{q}^2} {\mathbf{p}^2}}{6 {m_1}^2}+\frac{5 {\mathbf{q}^2} {\mathbf{p}^2}}{6
   {m_1}^2}+\frac{{k_v}^2 {\mathbf{q}^2} {\mathbf{p}^2}}{6 {m_2}^2}-\frac{7 {k_v} {\mathbf{q}^2} {\mathbf{p}^2}}{6 {m_2}^2}\nonumber\\&&-\frac{{E}^2 {\mathbf{q}^2} {\mathbf{p}^2}}{12
   {m_1}^2 {m_2}^2}-\frac{{E}^2 {k_v}^2 {\mathbf{q}^2} {\mathbf{p}^2}}{12 {m_1}^2 {m_2}^2}-\frac{{E}^2 {k_v} {\mathbf{q}^2} {\mathbf{p}^2}}{6 {m_1}^2
   {m_2}^2}+\frac{5 {\mathbf{q}^2} {\mathbf{p}^2}}{6 {m_2}^2}+\frac{25 {E}^2 {\mathbf{p}^2}}{12 {m_1}^2}-\frac{5 {E}^2 {k_v}^2 {\mathbf{p}^2}}{6 {m_1}^2}-\frac{5
   {E}^2 {k_v} {\mathbf{p}^2}}{12 {m_1}^2}\nonumber\\&&+\frac{25 {E}^2 {\mathbf{p}^2}}{12 {m_2}^2}-\frac{5 {E}^2 {k_v}^2 {\mathbf{p}^2}}{6 {m_2}^2}-\frac{5 {E}^2
   {k_v} {\mathbf{p}^2}}{12 {m_2}^2}+\frac{5 {\mathbf{p}^2}}{3}+5 {E}^2-20 {p_0}^2-\frac{{k_v}^2 {p_0}^2 {\mathbf{p}\cdot\mathbf{q}}^2}{{m_1}^2 {m_2}^2}-\frac{2
   {k_v} {p_0}^2 {\mathbf{p}\cdot\mathbf{q}}^2}{{m_1}^2 {m_2}^2}\nonumber\\&&-\frac{{p_0}^2 {\mathbf{p}\cdot\mathbf{q}}^2}{{m_1}^2 {m_2}^2}+\frac{17 {k_v}^2 {\mathbf{p}\cdot\mathbf{q}}^2}{6
   {m_1}^2}+\frac{{k_v} {\mathbf{p}\cdot\mathbf{q}}^2}{6 {m_1}^2}-\frac{5 {\mathbf{p}\cdot\mathbf{q}}^2}{6 {m_1}^2}+\frac{17 {k_v}^2 {\mathbf{p}\cdot\mathbf{q}}^2}{6 {m_2}^2}+\frac{{k_v}
   {\mathbf{p}\cdot\mathbf{q}}^2}{6 {m_2}^2}+\frac{{E}^2 {\mathbf{p}\cdot\mathbf{q}}^2}{4 {m_1}^2 {m_2}^2}\nonumber\\&&+\frac{{E}^2 {k_v}^2 {\mathbf{p}\cdot\mathbf{q}}^2}{4 {m_1}^2
   {m_2}^2}+\frac{{E}^2 {k_v} {\mathbf{p}\cdot\mathbf{q}}^2}{2 {m_1}^2 {m_2}^2}-\frac{5 {\mathbf{p}\cdot\mathbf{q}}^2}{6 {m_2}^2}+\frac{20 {k_v}^2 {\mathbf{p}\cdot\mathbf{q}}}{3}-\frac{10
   {k_v}^2 {p_0}^2 {\mathbf{p}\cdot\mathbf{q}}}{3 {m_1}^2}+\frac{5 {k_v} {p_0}^2 {\mathbf{p}\cdot\mathbf{q}}}{3 {m_1}^2}\nonumber\\&&+\frac{10 {p_0}^2 {\mathbf{p}\cdot\mathbf{q}}}{3 {m_1}^2}-\frac{10
   {k_v}^2 {p_0}^2 {\mathbf{p}\cdot\mathbf{q}}}{3 {m_2}^2}+\frac{5 {k_v} {p_0}^2 {\mathbf{p}\cdot\mathbf{q}}}{3 {m_2}^2}+\frac{10 {p_0}^2 {\mathbf{p}\cdot\mathbf{q}}}{3 {m_2}^2}-\frac{20
   {k_v} {\mathbf{p}\cdot\mathbf{q}}}{3}-\frac{10 {E} {k_v}^2 {p_0} {\mathbf{p}\cdot\mathbf{q}}}{3 {m_1}^2}\nonumber\\&&-\frac{5 {E} {k_v} {p_0} {\mathbf{p}\cdot\mathbf{q}}}{3
   {m_1}^2}+\frac{10 {E} {k_v}^2 {p_0} {\mathbf{p}\cdot\mathbf{q}}}{3 {m_2}^2}+\frac{5 {E} {k_v} {p_0} {\mathbf{p}\cdot\mathbf{q}}}{3 {m_2}^2}-\frac{5 {E}^2
   {\mathbf{p}\cdot\mathbf{q}}}{6 {m_1}^2}-\frac{5 {E}^2 {k_v}^2 {\mathbf{p}\cdot\mathbf{q}}}{6 {m_1}^2}-\frac{5 {E}^2 {k_v} {\mathbf{p}\cdot\mathbf{q}}}{4 {m_1}^2}\nonumber\\&&-\frac{5 {E}^2
   {\mathbf{p}\cdot\mathbf{q}}}{6 {m_2}^2}-\frac{5 {E}^2 {k_v}^2 {\mathbf{p}\cdot\mathbf{q}}}{6 {m_2}^2}-\frac{5 {E}^2 {k_v} {\mathbf{p}\cdot\mathbf{q}}}{4 {m_2}^2}+\frac{20
   {\mathbf{p}\cdot\mathbf{q}}}{3}+\frac{10 {k_v}^2 {\mathbf{q}^2}}{3}-\frac{10 {k_v} {\mathbf{q}^2}}{3}+\frac{5 {\mathbf{q}^2}}{3}
 \end{eqnarray}

 \begin{eqnarray}
 C_{\mathcal{T}2}=\frac{2 {\mathbf{p}^4}}{3 {m_1}^2 {m_2}^2}+\frac{5 {\mathbf{p}^2}}{3 {m_1}^2}+\frac{5 {\mathbf{p}^2}}{3 {m_2}^2}+5
  \end{eqnarray}

\end{document}